\documentstyle[12pt]{article}
\newcommand{\be}{\begin{equation}} \newcommand{\ee}{\end{equation}} 
\newcommand{\bea}{\begin{eqnarray}}\newcommand{\eea}{\end{eqnarray}}

\begin{document}
\begin{titlepage}
\title {Surface Tension at Finite Temperature in the MIT Bag Model} 
\vskip 3.0 cm 
\author{ {\bf Lina Paria}$^{1}$, {\bf M. G. Mustafa}$^{2, 3}$ and
{ \bf Afsar Abbas}$^{1}$  \\
 $ ^{1}$Institute of Physics, Sachivalaya Marg, \\
Bhubaneswar-751005, India. \\
$ ^{2} $ Institut fuer Theoretische Physik, \\   
Universitaet Giessen, Heinrich-Buff-Ring 16, \\                            
35392 Giessen, Germany.}
\footnotetext[1]{e-mail: \ lina/afsar@iopb.res.in} 
\footnotetext[2]{e-mail: \ munshi-golam.mustafa@theo.physik.uni-giessen.de} 
\footnotetext[3]{Alexander von Humboldt Fellow 
and on Leave from Saha Institute of Nuclear
Physics, \\ 1/AF Bidhan Nagar, Calcutta - 64, India}
\date{}
\maketitle
\thispagestyle{empty}
\begin{abstract}
At $ T = 0 $ the surface tension $ \sigma ^{1/3} $ in the MIT bag
model for a single hadron is known to be negligible as 
compared to the bag pressure
$ B^{1/4}$. We show that at finite temperature it has a substantial
value of 50 - 70 MeV which also differ from hadron to hadron. 
We also find that the dynamics of the
Quark-Gluon Plasma is such that the creation of hybrids $(s\bar{s}g)$
with massive quarks will predominate over the creation of $ (s\bar{s}) $
mesons. 
\end{abstract}
\end{titlepage}
\eject

\eject

\newpage

Recently a possible signature of Quark-Gluon Plasma (QGP) has been
proposed by two of us \cite{abba97}. We showed that the QGP 
dynamics is such that
during hadronisation the  creation of hybrids will predominate over
the creation of mesons. These hybrids will manifest  themselves
by significantly modifying the photonic signals of QGP. That
calculation \cite{abba97} was done for 2-flavour massless quarks.
One may ask what happens if one takes the massive quarks ?
In this paper we study this problem and look at the $(s\bar{s}g)$
hybrid with respect to the $(s\bar{s})$ mesons at finite
temperature and explore it's significance for QGP.

Recently it has been pointed out by Swanson \cite{swan98} that the hybrids may 
in fact have intrinsic connection with the concept of quark confinement. 
If this is indeed true then the study of hybrid should be more crucial than
was realised earlier. This paper is a continuing venture in that
direction.

We also look at the surface tension at finite temperature in 
the MIT bag model for a single hadron. It is known \cite{farhi84} that 
at $ T = 0 $, the surface tension in the MIT bag model 
for a single hadron is negligible in
comparison to the bag pressure $ B^{1/4} $. (Note that
what is called ``intrinsic" surface tension in ref.\cite{farhi84} is
what we have here and shall call it simply ``surface tension").
However for a massive quark (like s-quark) at finite 
temperature the surface tension $ \sigma^{1/3}$ may be non-zero \cite{taka87}.
Recently some work has been done on the surface tension both at
$T \ne 0 $ \cite{mard91} and at $ T = 0 $ \cite{mad94, christ97}.
A mass formula for a spherical lumps of three-flavour quark matter at $T = 0$
was derived within the MIT bag model, taking into account bulk, surface
and curvature contributions where an ansatz is provided for
the curvature contribution to the density of states for massive
quarks \cite{mad94}. At finite temperature near the
confinement phase transition, the free energy of finite droplets
of QGP and of finite hadronic bubbles in an bulk plasma has been
calculated by Mardor $\&$ Svetitsky \cite{mard91}. They have
shown that the curvature term in the free energy proportional
to the radius of the droplet or bubble is more important than the
contribution of the surface tension, proportional to the radius squared.
But it is not clear what its exact value in the MIT bag should be. 
Here we find that the (intrinsic) surface 
tension $\sigma^{1/3}$ at finite temperature has a substantial
value and is in the range 50-70 MeV which also depends upon the
structure of hadrons. Hence the same (intrinsic) surface
tension which is zero at $ T = 0 $  is shown to have a substantial non-zero
value at finite temperature.

\section{Formalism : }

To study the thermodynamic properties of a hadron (assumed to be in
contact with the heat bath mimicking the hot central zone in relativistic
heavy ion collision), a bag is treated like a many-body system 
at a given temperature. The thermodynamic treatment of a hot 
hadronic bag has been justified by several authors over the 
years \cite{taka87, kars80, kapu81, jenn82, jd89}.
Which pointed out that at a high temperature a large number of
$q \bar{q}$ pairs can be created from the negative energy sea, and
so at least in principle it is not a system of only two or three particles.
We consider a system of non-interacting quarks, anti-quarks and gluons
placed in a heat bath with which it can exchange the energy and
the particle numbers, so that one can use the grand canonical ensemble
formalism. The quarks and gluons are confined in an MIT bag
\cite{chod74}. We consider the system of single quark flavor (s-quark)
with mass $ m_s = 150 $ MeV where the single particle energies 
are obtained by solving the equation of motion with linearised boundary
condition \cite{deg75}. The single particle energies for the massive quarks
and massless gluons are in units of ${\hbar c} /R $, where the radius of the
bag $R$ is included.

In a statistical approach, the grand canonical partition function is
given by
\be
Z_G \ = \ Z_{\rm {vac}} \ Z_{s} \ Z_{\bar{s}} \ Z_{g}   \label{surf1}
\ee

\noindent where $ Z_{\rm {vac}} $ takes care of the
temperature $ T \rightarrow 0 $ limit, and we consider this as
\be
-T \ln Z_{\rm {vac}} \ = \ BV \ + \ C/R   \label{surf2}
\ee
\noindent with $ BV $ being the volume energy of the bag and $ C/R$ as
the Casimir energy which is $T$ independent
with the value of C given in ref.\cite{abba97, jd89, mun93}.
$ Z_{s} $ and $Z_{\bar s}$
refers to the partition function for the s-quark and s-antiquark, where
$Z_{g} $ refers to the gluonic part.

The logarithm of the partition function for the system of s-quark, 
anti-quark and gluon with the chemical potential 
$ \mu_{s(\bar s)}$ for quark(anti-quark)
is given by
\bea
 \ln Z \ & =  & \  {\sum_i} \ \ln \ {\Big (} \ 1  \ + 
	\ e^{- {({\epsilon_i}^{s} - \mu_s)/T}} {\Big)} 
 \ + \  {\sum_i}  \  \ln \ {\Big (} \ 1  \ +  
 \ e^{- {({\epsilon_i}^{\bar s} - \mu_{\bar s})/T}} {\Big)}  \\ \nonumber
\ & - & \  {\sum_i}  \ \ln \ {\Big (} \ 1  \ -  
 \ e^{- {{{\epsilon_i}^{g}}/T}} {\Big)}             \label {surf3}
\eea

\noindent where $\epsilon_i^{s (\bar s)}$ and $\epsilon_i^g$ are the
quark (anti-quark) and gluon single particle energies in the bag
respectively.

The number of quark and gluon is given by
\be
N_s  \ = \ {\sum_i} \ {1 \Big{/}{\Big(} \ e^{({\epsilon_{i}}^{s}- \mu_s )/T} 
\ + \ 1{\Big)}}   \label{surf4}
\ee
and
\be
N_g \ = \ {\sum_i} \ {1 \Big{/}{\Big(} \ e^{{\epsilon_{i}}^{g} /T} 
\ - \ 1{\Big)}}     \label{surf5}
\ee
\noindent Whereas the energy and the free energy 
of the quark gluon system is given by

\be
E(T, R) \  =  \ T^2 \ {\Big(} {\partial \ \ln Z \over
{\partial T}} {\Big)} \ + \mu_{s(\bar s)} N_{s(\bar s)} 
\ + \ BV \ + \ C/R     \label{surf6}
\ee

\be
F(T, R) \ = \ -T \ \ln Z \ + \ \mu_{s(\bar s)} N_{s(\bar s)} 
\ + \ BV \ + C/R     \label{surf7}
\ee
The pressure generated by the participant gas 
\be
P \ = \ - \ {\Big(} {\partial \ F(T, V) \over {\partial V}} 
{\Big)}_{T, \ \mu_s}       \label{surf8}
\ee
\noindent is balanced by the bag pressure constant $B$ leading to the
stability condition to the system.

\section {Results and discussion : }

The physical behaviour of a system at a finite temperature T is governed by
the properties of its free energy. We consider the radius of the bag $R$
to be a variational parameter in the free energy $F(T, R)$
to study the stability of the bag. The quarks and
gluons single particle energies are in units of ${\hbar c}/R$, 
so for a fixed value of $T$, we vary $R$ while adjusting the
chemical potential and ensuring that we have one s-quark in the system. 
For this temperature
$T$ and for the range of $ R$ considered, the gluon number $N_g$ is calculated
from Eq.(\ref{surf5}).
If this value is less than one (which is true for the low temperatures), we
go to higher temperature such that we have one gluon.
Now at this temperature $T$, we calculate the free energy of the $(s \bar s g)$
system with different values of $R$ and see that there is a radius $R$ of
the bag for which there is one gluon attaches to the $(s\bar s)$ system making 
it hybrid $(s \bar s g)$ and also at which point the free energy of the 
hybrid is minimum.

For $B^{1/4} = 200$ MeV, we calculate the free energy of the $(s \bar s g)$
system for various $R$ values at a fixed temperature $T$ at which $F$ is
minimum at some value of $R$ where one gluon attaches to the $(s \bar s)$
system making it hybrid $(s \bar s g)$ at that $T$.

At the same temperature, the free energy of 
the pure mesonic system $(s\bar{s})$ has been calculated as a function
of radius $R$. This variations of free energies for hybrid $(s\bar{s}g)$
and meson $(s\bar{s})$ is displayed at Fig. 1 and the corresponding
energy (mass) at the minimum free energy condition is given in Table 1.

From Fig. 1 we see that at the temperature $ T = 156 $ MeV ; the
hybrid $(s\bar{s}g)$ is more stable (less free energy) than the
meson ($s\bar{s})$, whereas from Table 1 we see that the energy (mass)
of the hybrid ($s\bar{s}g)$ is less than twice the mass of the
meson $(s\bar{s})$ hence forbidding the  decay of the strange
hybrid $(s\bar{s}g)$ into a pair of mesons $(s\bar{s})$.
So although $(s\bar{s}g)$ hybrid decay into a pair of $(s\bar{s})$
mesons at $ T =  0 $ MeV \cite{chan83, clos95}, the strange
hybrid ($s\bar{s}g)$ does not decay into ($s\bar{s})$ pair at finite T.
However unlike the case of non-strange hybrids, as
$s \bar{s} g \rightarrow s\bar{u} \ + \ \bar{s} u $ is possible and
hence strange hybrid may decay through strong interaction. Earlier we had
shown \cite{abba97} that the non-strange hybrids will leave a unique signature
in terms of suppression of photonic signals. The same may not be true
of strange hybrids.

Note that our picture of the hybrid at finite temperature is akin
to what is considered \cite{clos95} as gluonic excitation of meson.
So what may have been a meson at $ T = 0 $ converts into a hybrid
at high temperature due to the gluonic excitation arising therein.

For the case of strange hybrid $(s\bar{s}g)$ we see that
during the hadronisation of thermalised QGP,  the formation
of hybrid is more favourable than that of the meson at the
same temperature. So the mixed phase in addition to QGP and mesons
will contain massive hybrids as well and thereby reduce 
the life-time of the mixed phase.
Hence by studying the non-strange hybrid \cite{abba97}
as well as the strange hybrid we found that the hybrid formation
is favoured over the meson formation during the hadronisation 
of the QGP.

For a system of quarks and gluons confined in a bag of finite
size, the finite size corrections are significant.
The finite size corrections can be incorporated in two different ways.
The first method is to take the sum over the discrete single
particle states of quarks and gluons for computing the thermodynamic
quantities \cite{jd89} and also as we have done here. The second 
method is to replace the sum over
discrete states by the integral with single particle density of states
including finite size corrections \cite{mun93}. For reasonably
dense system (as we have here) these two methods are equivalent to each
other \cite{farhi84, jd89, mun93}. We shall use this equivalence 
to calculate the (intrinsic) surface tension of the single hadron 
at finite temperature.

For the interior of the sphere with radius R, the density of
state becomes \cite{farhi84, mun93}

\be
\rho(k) \ = \ { 2 \over {3 \pi }} \ R^3 \ k^2 \ + \ C_1 \ 4 \
\pi \ R^2 \ k \ + \ C_2 \ 8 \pi \ R      \label{surf10}
\ee

\noindent

Since the surface area term vanishes in $ \rho(k) $ for the
massless particles, one gets the mass (energy) of the system
of massless quarks, anti-quarks and gluons at the equilibrium
condition as $ M = 4 B V $. Where B is the usual bag pressure
constant. But for the system of massive quarks the surface
term also does contribute. So for the strange quark system
we add the surface energy term $ 4 \pi R^2 \sigma $ in the
energy (Eq.(\ref{surf6})) and free energy 
expression (Eq.(\ref{surf7})). In these expressions we replace
the discrete sum over single particle states by the integral with
density of states containing the volume and curvature term. Now
applying the pressure balance condition as Eq.(\ref{surf8})
for stability , 
the mass (energy) of the system 
\cite{taka87, mun93} at equilibrium
(where the total free energy is minimum) is given by

\be
M \ = \ 4 \ B \ V \ + \ 12 \ \pi \ R^2 \ \sigma     \label{surf11}
\ee

\noindent

Where we parametrize $ \sigma $ as the coefficient of the surface energy
and is known as surface tension.

In our system of strange hybrid $(s\bar{s}g)$ at finite temperature,
we get the mass of the hybrid at equilibrium (minimum free energy
position) which should be equal to the expression (\ref{surf11}) as our
system has massive quarks for which surface energy does contribute.
Hence we get,

\be
\sigma^{1/3} \ = \ {{\Big [ } \ {{ M - 4 B V } \over 
{12 \pi R^2 }} \ {\Big ] }}^{1/3}  \label{surf12}
\ee

\noindent

With $ B^{1/4}  =  200$  MeV, for the hybrid of 
$ R  =  0.943 $ fm,
and $ M = 3.086 $ GeV, we get the value of the surface
tension as $ \sigma^{1/3} = 57 $ MeV at
the temperature $ T = 156$ MeV. Whereas at  the same temperature
$ T = 156 $ MeV, the surface tension for the meson $ (s\bar{s})$
is obtained as $ \sigma^{1/3} = 63 $ MeV.

Similarly for $ B^{1/4} = 250$ MeV, 
the $ (s\bar{s}g) $ hybrid is formed at the temperature
$ T = 195 $ MeV  with  radius $ R = 0.758 $ fm,
and mass $ M = 3.87 $ GeV, so the value of 
$ \sigma^{1/3} = 66 $ MeV  whereas the value of $ \sigma^{1/3}$
for ($s\bar{s}$) meson with radius $ R = 0.647 $ fm and mass
$ M_{(s\bar{s})} = 2.459 $ GeV is $ 72 $ MeV.
For $ B^{1/4} = 150 $ MeV, we find $ \sigma^{1/3} (s \bar{s} g) = 48 $
MeV and $ \sigma^{1/3} (s \bar{s}) = 53 $ MeV.
Hence at the same temperature and bag pressure, the value
of surface tension $ (\sigma^{1/3}) $  is larger for pure meson
than for the hybrid. 
It is expected that different hadrons should have different
surface tensions depending upon their constituent objects \cite{berg89}.
We would like to point out the power of our method is that
it allows us to extract surface tension of objects what have
gluonic degree as well.

We have looked at the $ (s \bar{s}) $ system for various temperatures
$ T = 5, 10, 50, 100, 150 $ MeV for $ B^{1/4} = 200 $ MeV and
found $ \sigma^{ 1/3} $ to be almost steady with a slight decrease
from 68 MeV to 63 MeV as the temperature increases. However
one should be warned that our method is justified for the cases
where the continuum approximation is valid, that is at higher
temperatures. Also for $ B^{1/4} = 200 $ MeV, $\sigma^{1/3}(s \bar{s}g) $
goes from 39 MeV to 75 MeV and $ \sigma^{1/3}(s\bar{s}) $ goes from 
43 MeV to 82 MeV as $m_s $ goes from 50 MeV to 300 MeV. In the same
range the temperature changes very slightly from 155 MeV to 157 MeV.
In ref.\cite{berg89} Berger finds that the dynamical surface tension decreases
for heavy strange quark masses. This was found to be true
as the number density of the strange quarks in their
system decreased as the mass was increased. However in our calculation here
we find that the radius of the hybrid and the meson are almost stationary
as the strange quark mass increases and also that we fix the number of
quark and antiquark. Hence the number density in our
calculation does not change as the strange quark mass increases.
Therefore the increase in strange quark mass is directly reflected
in terms of an increase of the corresponding (intrinsic) surface
tension as we find here.

Note that our surface tension here is the intrinsic surface
tension \cite{farhi84}.  This is different from the surface tension as
obtained in the lattice calculations. The latter is that of the hadron
phase and the QGP phase as separated by an interface at the
phase transition. These values have gone through extreme evolutions.
A prior one expects $ {\sigma/{T_c}^3} $ to be of the order of 
unity \cite{kaja90}. 
However when calculations were done it
was found \cite{kaja90} to just have a value of $ \sim 0.24 $. Latter better
lattice calculations brought it down to be $ \sim 0.027 $ 
\cite{brow92, iwas94}.
It is amusing to note that these values are not very different from 
our results in this paper.

\vskip 6.0 cm
Acknowledgement : MGM gratefully acknowledges the financial support
from the Alexander von Humboldt Foundation, Bonn, Germany.


\vfill

\newpage

\noindent {\centerline {\bf CAPTIONS}}
\noindent {\bf Table 1} \\
The equilibrium radius $(R)$ 
and the energy (mass) $(E)$ of the hybrid $ (s\bar{s}g)$ and pure mesonic
system $ (s\bar{s})$ is given for bag constant $ B^{1/4} = 200 $ MeV.

\vskip 7.0 cm

\noindent {\bf Figure 1.} \\
The variation of the free energy of the hybrid $(s\bar{s}g)$
and the pure mesonic systems $(s\bar{s})$ with radius $R$ are displayed
at a particular temperature $ T = 156 $ MeV and
 $ B^{1/4} = 200 $ MeV.

\vfill

\newpage

\begin{table}
\centerline {\bf Table 1} 
\vskip 0.2 in
\begin{center}
\begin{tabular}{|c|c|c|c|}
\hline
System & T(MeV) & R(fm) & E(GeV) \\
\hline
($s\bar{s}g$) & 156 & 0.943 & 3.086 \\
\hline
($s\bar{s}$) & 156 & 0.808 & 1.996   \\ 
\hline
\end{tabular}
\end{center}
\end{table}
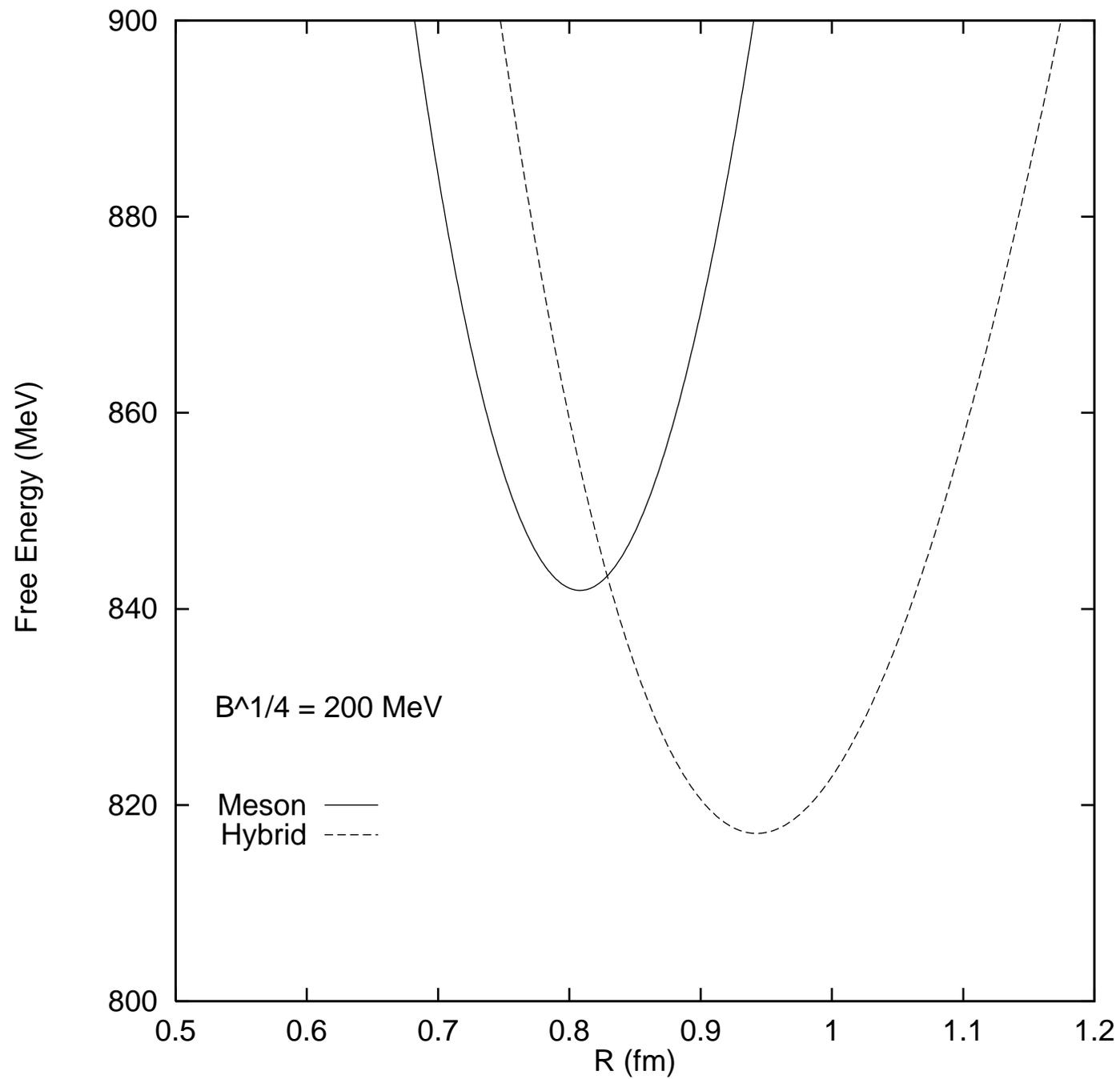
\end{document}